\documentclass[aps,prd,reprint,nofootinbib,superscriptaddress]{revtex4-2}
\usepackage{bm}
\usepackage{amsmath}
\usepackage{amssymb}
\usepackage{amsfonts}
\usepackage{color,xcolor}
\usepackage{graphicx}
\usepackage{float}
\usepackage{subfigure}
\usepackage{hyperref}
\usepackage{comment}
\usepackage{ulem}
\usepackage{CJK}
\usepackage{dcolumn}
\hypersetup{colorlinks=true,linkcolor=blue,citecolor=red}

\newcommand{\red}{\textcolor[rgb]{1.00,0.00,0.00}}

\begin{document}

\title{Black holes with scalar hair: Extending from and beyond the Schwarzschild solution}

\author{Xi-Jing Wang}
\email{xijingwang@yzu.edu.cn}
\affiliation{Center for Gravitation and Cosmology, College of Physical Science and Technology, Yangzhou University, Yangzhou, 225009, China}

\author{Guoyang Fu}
\email{FuguoyangEDU@163.com}
\affiliation{Center for Gravitation and Cosmology, College of Physical Science and Technology, Yangzhou University, Yangzhou, 225009, China}

\author{Peng Liu}
\email{phylp@email.jnu.edu.cn}
\affiliation{Department of Physics and Siyuan Laboratory, Jinan University, Guangzhou 510632, P.R. China}

\author{Xiao-Mei Kuang}
\email{xmeikuang@yzu.edu.cn}
\affiliation{Center for Gravitation and Cosmology, College of Physical Science and Technology, Yangzhou University, Yangzhou, 225009, China}

\author{Bin Wang}
\email{wang$_$b@sjtu.edu.cn}
\affiliation{Center for Gravitation and Cosmology, College of Physical Science and Technology, Yangzhou University, Yangzhou, 225009, China}
\affiliation{Shanghai Frontier Science Center for Gravitational Wave Detection, Shanghai Jiao Tong University, Shanghai 200240, China}

\author{Jian-Pin Wu}
\email{jianpinwu@yzu.edu.cn, corresponding author}
\affiliation{Center for Gravitation and Cosmology, College of Physical Science and Technology, Yangzhou University, Yangzhou, 225009, China}

\begin{abstract}

We construct novel scalarized black hole (BH) solutions beyond the general relativity (GR) framework. These scalarized BH solutions are extended from the Schwarzschild one and the non-Schwarzschild one in the pure Einstein-Weyl gravity. By studying the BH entropy and free energy, we demonstrate that the scalarized BH extending from the Schwarzschild one exhibits thermodynamically preferred. We obtain these novel solutions by directly solving the full fourth-order equations of motion. This narrows the problematic solution space obtained by commonly adopted second-order reduction to physically valid spaces. Our findings also unveil the evasion of the no-hair theorem within the realm of higher-derivative gravity.

\end{abstract}

\maketitle
\section{Introduction}

The direct detection of gravitational waves (GWs) \cite{LIGOScientific:2016aoc, LIGOScientific:2016lio, LIGOScientific:2016sjg} and the images of black holes (BHs) \cite{EventHorizonTelescope:2019dse, EventHorizonTelescope:2019ths, EventHorizonTelescope:2022wkp, EventHorizonTelescope:2022xqj} has heralded a new era in physics, providing powerful tools for exploring the realm of strong gravity. These breakthroughs have also opened up new avenues for investigating the potential existence of new fundamental scalar fields that may leave distinctive imprints on BHs \cite{Berti:2015itd,Herdeiro:2015waa}. However, the no-hair theorem in general relativity (GR) presents a significant obstacle, as it precludes the existence of stationary, asymptotically flat BH solutions with scalar hairs \cite{Israel:1967wq,Carter:1971zc,Wald:1971iw,Ruffini:1971bza}. This theorem has been further extended to encompass other theories of gravity, including scalar-tensor theories \cite{Hawking:1972qk,Bekenstein:1995un,Sotiriou:2011dz} and $f(R)$ gravity \cite{Canate:2017bao,Sultana:2018fkw}.

However, it is possible to bypass the no-hair theorem by violating some of their assumptions \cite{Cardoso:2013fwa,Cardoso:2013opa,Babichev:2013cya,Herdeiro:2014goa,Sotiriou:2015pka,Herdeiro:2015waa}. One of the ways is to introduce a scalar field that is non-minimally coupled to the Gauss-Bonnet (GB) invariant. This coupling leads to the emergence of scalarized BHs, which are extended from the Schwarzschild BH in GR \cite{Doneva:2017bvd,Silva:2017uqg,Antoniou:2017acq}. The presence of scalarized BHs enables the detection of scalars in these theories by observing the distinctive imprint they leave when excited \cite{Barausse:2008xv,Arvanitaki:2009fg}, and serves as a valuable criterion for singling out particularly intriguing theories.

The GB invariant, despite involving higher derivatives, exhibits relative simplicity with its equations of motion (EOMs) remaining second order. However, it remains uncertain whether the inclusion of more extensive higher derivatives will also lead to scalarization. This consideration holds significant meaning as it has the potential to extend the existence of scalar hair to a significantly broader range of scenarios.

Motivated by these considerations, we focus our attention on a novel geometric invariant $C^2$ which is defined as $C^2\equiv C_{\mu\nu\rho\sigma} C^{\mu\nu\rho\sigma}$ coupled to a scalar field $\varphi$. The action we consider is
\begin{equation}
	S=\frac{1}{16\pi} \int d^4 x \sqrt{-g} \left[ R-2 \nabla_\mu \varphi\nabla^\mu \varphi-\alpha F(\varphi)C^2 \right]\,.\label{action}
\end{equation}
Here $\alpha$ represents the coupling constant with dimensions of $[length]^2$, while $F(\varphi)$ is a coupling function that solely depends on the scalar field $\varphi$.

When $\varphi=0$ and $F(\varphi)=1$, the theory \eqref{action} reduces to the pure Einstein-Weyl theory \cite{Lu:2015cqa}:
\begin{equation}
	S=\frac{1}{16\pi} \int d^4 x \sqrt{-g} \left(R-\alpha C_{\mu\nu\rho\sigma} C^{\mu\nu\rho\sigma} \right)\,.\label{action1}
\end{equation}
This theory is an extension of Einstein gravity, incorporating higher-order derivative terms that arise in string theory and other effective theories, and has been proven to be renormalizable in four-dimensional spacetime \cite{Stelle:1976gc}. However, it is worth noting that this theory also possesses intrinsic significance and relevance in its own right. Additionally, as demonstrated in \cite{Lu:2015cqa}, this theory describes a system with a massive spin-2 mode with a squared mass of $m^2=1/(2\alpha)$, in addition to the massless spin-2 graviton. Therefore, $\alpha$ is generally a positive parameter. Moreover, this theory encompasses not only the conventional Schwarzschild solution but also another branch of BH solution, referred to as the non-Schwarzschild solution, which significantly deviates from the Schwarzschild geometry \cite{Lu:2015cqa}. Although novel non-Schwarzschild solutions can be obtained within this theory, it should be noted that these novel solutions are not thermodynamically favored. An interesting aspect to consider is whether the scalarized solutions, when considering scalarization, would respond to the two branches of solutions and, more importantly, be more thermodynamically favored.

These theories lead to fourth-order differential equations for the metric functions, which can be challenging to solve directly. Previous studies have shown it is possible to reduce these equations to second-order by eliminating the higher-derivative terms \cite{Lu:2015cqa,Huang:2022urr}, which may seem to simplify the analysis. However, we argue that the full fourth-order equations contain important information that could be lost in the reduction process. In particular, the higher-derivative nature of the theory points to new dynamics that depend intrinsically on the fourth-order scale set by $\alpha$. Reducing the order of the equations prematurely may ignore these novel effects. Therefore, in this work, we tackle the full fourth-order differential equations directly using numerical methods. This allows us to reveal the true structure of the higher-derivative theory.

This paper aims to unveil the existence of scalarized black holes, their profiles, and thermodynamic stability in Einstein-scalar-Weyl gravity \eqref{action} by directly solving the fourth-order EOMs.

\section{Scalarized black holes}\label{S-NR}

We are interested in a geometry that is static, spherically symmetric, and asymptotically flat, described by the following metric ansatz:
\begin{eqnarray}
	&&
	ds^2 = -h(z)dt^2 + \frac{1}{z^4f(z)}dz^2 + \frac{1}{z^2}(d\theta^2 + \sin^2\theta d\phi^2),\nonumber
	\
	\\
	&&
	h(z)=(1-z)U_1(z)\,,\,\,\,\,\,\,\,\,\, f(z)=(1-z)U_2(z)\,.
	\label{ansatz}
\end{eqnarray}
We assume that the scalar field $\varphi$ depends solely on the radial coordinate $z$, i.e., $\varphi=\varphi(z)$. In our convention, the event horizon is located at $z=1$ and the infinite boundary at $z=0$. In addition, we select a particular scalar coupling function $F(\varphi)=e^{-\beta \varphi}$. When $\beta=0$, the BH solutions will reduce to those in pure Einstein-Weyl gravity \cite{Lu:2015cqa}. Due to the symmetry of the theory under the transformations $\beta\rightarrow-\beta$ and $\phi\rightarrow-\phi$, we solely consider $\beta>0$.

The EOMs resulting from the action \eqref{action} with the ansatz \eqref{ansatz} involve fourth-order derivatives of the metric functions $h(z)$ and $f(z)$. Following a similar process as \cite{Lu:2015cqa,Huang:2022urr}, it is possible to reduce these to second-order by eliminating higher-derivative terms, which seems to simplify the analysis. However, just as argued in the introduction, any true solution must satisfy both the second-order and original fourth-order equations. The reduction process can discard key constraints, so solutions from just the reduced equations may be invalid\footnote{Despite utilizing the reduced second-order equations to find solutions in pure Einstein-Weyl gravity, we have validated the black hole solutions presented in \cite{Lu:2015cqa}, which satisfy both the second-order and fourth-order equations simultaneously. Hence, their findings are valid. However, the multiple solution cases demonstrated in the subsequent developments only comply with the reduced second-order equations and fail to satisfy the fourth-order equations.}. To obtain robust solutions reflecting the complete theory, we directly solve the fourth-order system using pseudo-spectral methods \cite{Horowitz:2012ky,Ling:2014saa}. Please see the appendix \ref{appendix:methods} for more details.

Near the infinity the asymptotic behaviors for the metric functions and scalar field read as
\begin{align}
	h(z) & = 1-2 M z+\cdots,\quad f(z)=1-2 M z+\cdots,\nonumber \\ \varphi(z)&=\varphi_{\infty}+D z+\cdots\,,
	\label{expanInf}
\end{align}
where we denote the mass of the BH as $M$, the asymptotic value of the scalar field as $\varphi_{\infty}$, and the scalar charge as $D$. At infinity, we impose the boundary conditions $U_1(0)=U_2(0)=1$, and at the horizon, without loss of generality, we set the scalar field $\varphi(1)=1$. Other boundary conditions can be fixed by the regularity.

Before proceeding, it is worth noting that we have conducted extensive numerical simulations. We have verified that the fourth-order solutions satisfy both equation systems, while second-order solutions generally fail to satisfy the fourth-order equations. By solving the full system, we not only respect the higher-derivative structure but also identify the true solutions that constitute a subset of the larger but incorrect solution space obtained through premature reduction. See appendix \red{\ref{appendix:validation}} for more details.

\subsection{Scalarized black hole extending from the Schwarzschild solution} \label{case1}

\begin{figure}[htbp]
	\centering
	\includegraphics[width=0.23\textwidth]{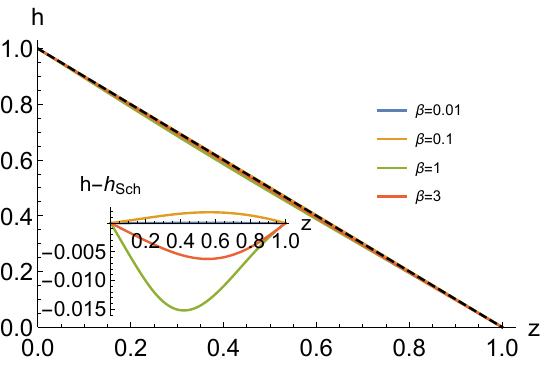}\hspace{2mm}
	\includegraphics[width=0.23\textwidth]{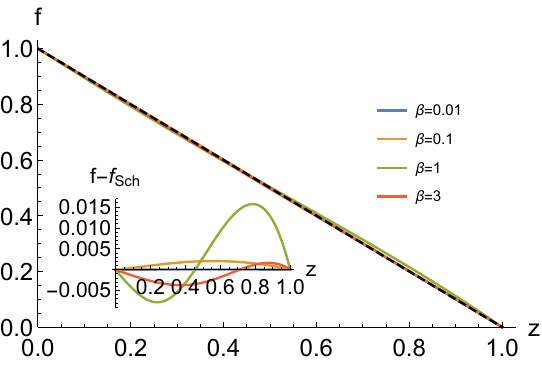}\hspace{2mm}
	\includegraphics[width=0.23\textwidth]{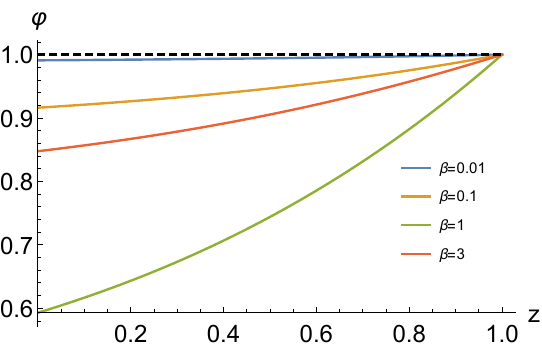}
	\caption{The metric fields $h$ and $f$, along with the scalar field $\varphi$ as functions of the radial coordinate $z$ for different parameter $\beta$. The black dashed curves are the Schwarzschild solution with trivial scalar field. The inset plots show the differences between the hairy BH and the Schwarzschild solution.}
	\label{case1-1}
\end{figure}

Throughout this paper, we set $\alpha=0.5$, as chosen in \cite{Lu:2015cqa}, while allowing $\beta$ to vary as a free parameter. We begin by examining the behavior of the metric fields $h$ and $f$, as well as the scalar field $\varphi$, as functions of the radial coordinate $z$ for different values of $\beta$. These plots are presented in Fig. \ref{case1-1}. The Schwarzschild solution with a trivial scalar field is represented by the black dashed lines. We can observe slight deviations between the hairy BH and the Schwarzschild solution, as highlighted in the inset plots.


Although the dilaton-like coupling function $F(\varphi)=e^{-\beta\varphi}$ only allows nontrivial solutions for the scalar field $\varphi$ according to the EOMs, we propose that as $\beta$ approaches zero, the hairy black hole gradually converges to the Schwarzschild solution with a vanishing scalar field.
This convergence is supported by the asymptotic behavior of the coupling function $F(\varphi)$, which tends to unity in the limit of $\beta \rightarrow 0$. Consequently, the Einstein-scalar-Weyl system \eqref{action} reduces to the pure Einstein-Weyl system. This result indicates the Einstein-scalar-Weyl theory can evade the no-hair theorem and permits the existence of a hairy black hole that seamlessly extends from the Schwarzschild solution within the framework of pure Einstein-Weyl gravity.

\begin{figure}[htbp]
	\centering
	\includegraphics[width=0.23\textwidth]{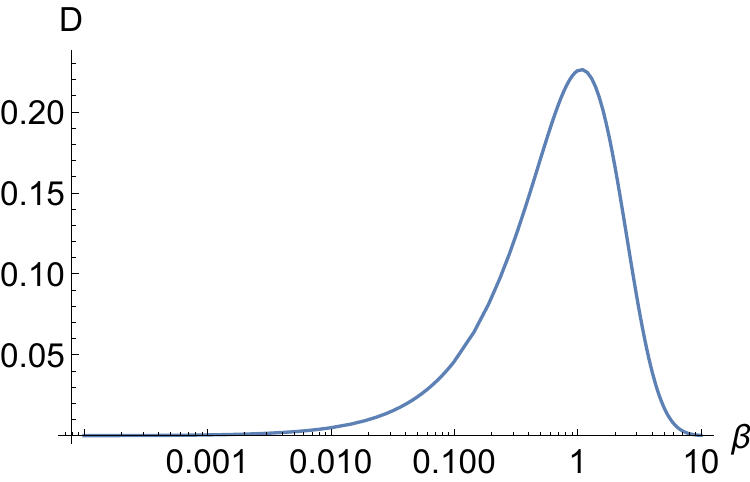}\hspace{2mm}
	\includegraphics[width=0.23\textwidth]{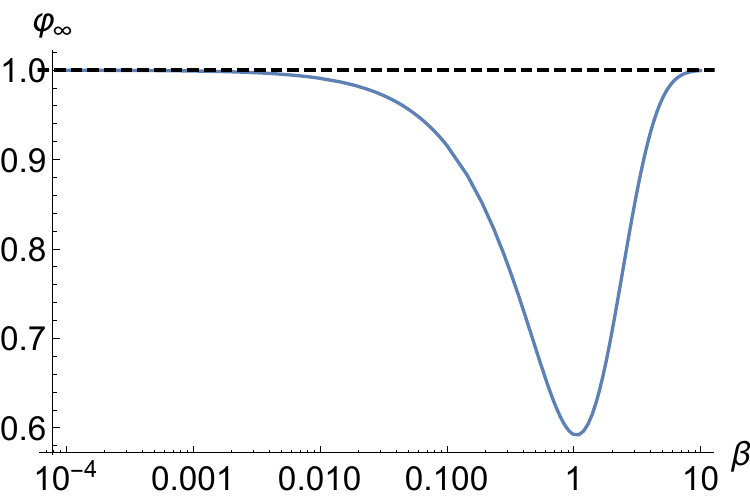}\hspace{2mm}
	\includegraphics[width=0.23\textwidth]{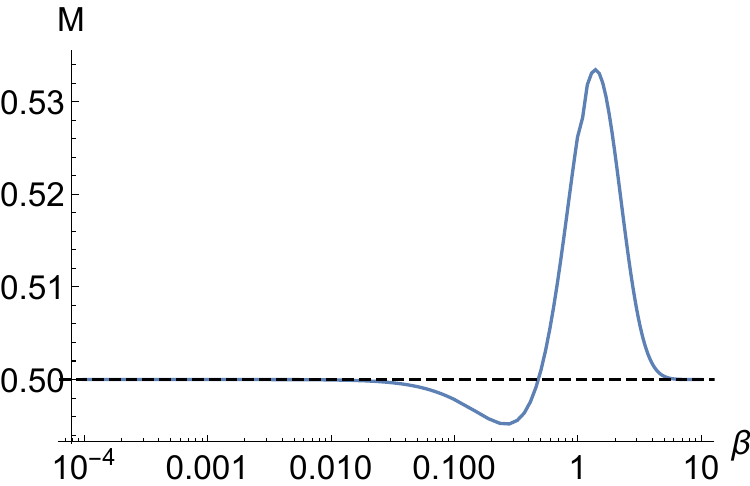}
	\caption{The scalar charge $D$, $\varphi_{\infty}$ and the BH mass $M$ as a function of $\beta$. The black dashed curves denote the case with trivial scalar hair.}
	\label{case1-2}
\end{figure}

We proceed by examining the behaviors of the scalar charge $D$, the asymptotic value of the scalar field denoted as $\varphi_{\infty}$, and the BH mass $M$ as we vary $\beta$. These results are displayed in Fig.\ref{case1-2}. Notably, the scalar charge exhibits an upward trend with increasing $\beta$, starting from zero, which corresponds to the Schwarzschild case. It reaches its maximum value when $\beta \approx 1$ and subsequently decreases, eventually approaching zero. In contrast, as $\beta$ increases, $\varphi_{\infty}$ exhibits an opposite trend in comparison to the scalar charge. The scalarization process results in the change of the BH mass, as depicted in Fig.\ref{case1-2}. It is observed that the BH mass initially decreases as $\beta$ increases from the mass of the Schwarzschild BH, then it begins to increase. The maximum value is reached at $\beta \approx 1$, after which it gradually decreases, eventually approaches the mass of the Schwarzschild BH.

It is worth noting that as $\beta$ approaches infinite, the system shifts from Einstein-scalar-Weyl gravity to Einstein-scalar gravity. This occurs because when $\beta$ approaches infinity, the coupling function $F(\varphi)$ approaches zero. Consequently, the impact of the Weyl term becomes negligible, and the BH solution reduces to the Schwarzschild solution described by pure Einstein gravity. In addition, we would like to point out that the scalar charge $D$ depends on the BH mass $M$. As a result, the hair in this scenario is secondary.

\subsection{Scalarized black hole beyond the Schwarzschild solution} \label{case2}

\begin{figure}[htbp]
	\centering
	\includegraphics[width=0.23\textwidth]{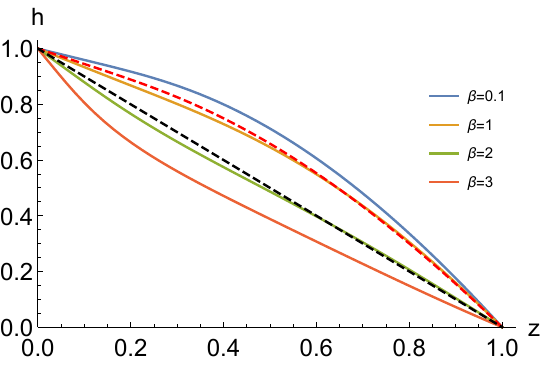}\hspace{2mm}\includegraphics[width=0.23\textwidth]{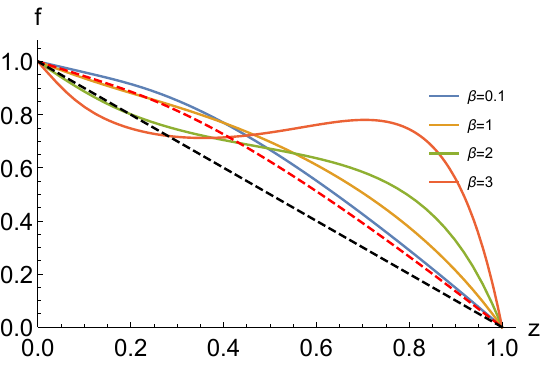}\hspace{2mm}
	\includegraphics[width=0.23\textwidth]{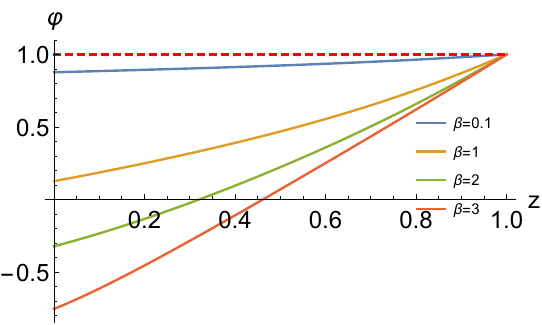}
	\caption{The metric ($h$ and $f$) and scalar ($\varphi$) functions  for different parameter $\beta$. The black dashed curves represent the Schwarzschild solution with a trivial scalar field, whereas the red dashed curves  depict the non-Schwarzschild solution with a trivial scalar field.}
	\label{case2-1}
\end{figure}

In this subsection, we address the emergence of a distinctive class of hairy BH solutions for the same parameters, which is shown in Fig.\ref{case2-1}. From this figure, it is evident that these solutions exhibit substantial deviations from both the Schwarzschild case (the black dashed curves in Fig.\ref{case2-1}) and the non-Schwarzschild case (the red dashed curves in Fig.\ref{case2-1}) as the parameter $\beta$ increases.
Remarkably, as $\beta$ approaches zero, the hairy BH converges to the non-Schwarzschild solution, indicating that this solution extends from the non-Schwarzschild BH in pure Einstein-Weyl gravity. It is different from the one we obtained in the previous subsection.

\begin{figure}[htbp]
	\centering
	\includegraphics[width=0.23\textwidth]{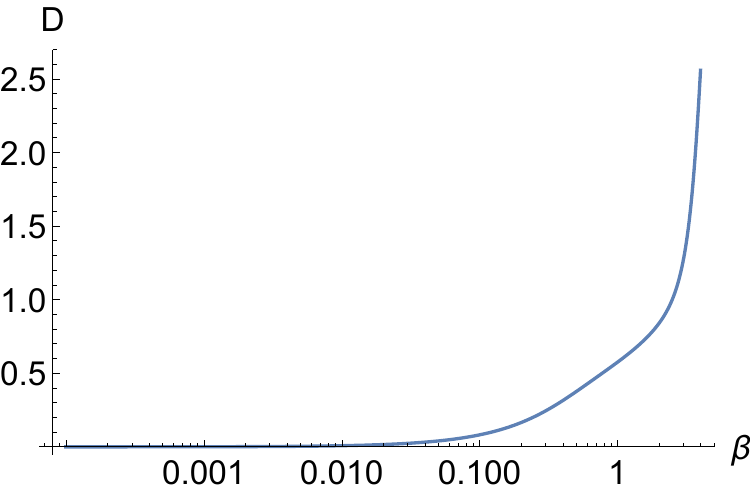}\hspace{2mm}
	\includegraphics[width=0.23\textwidth]{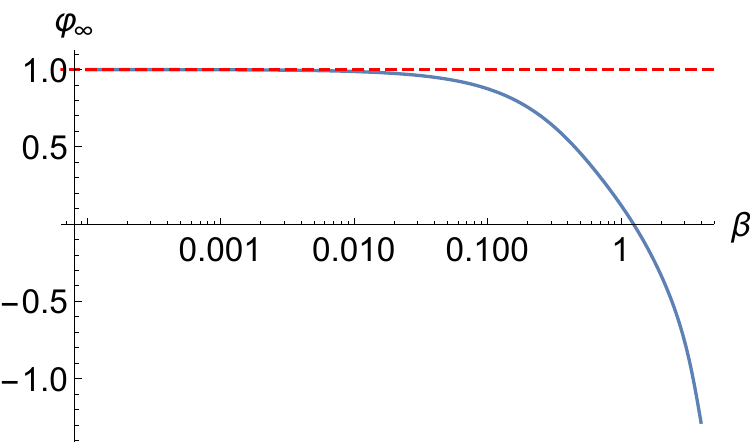}\hspace{2mm}
	\includegraphics[width=0.23\textwidth]{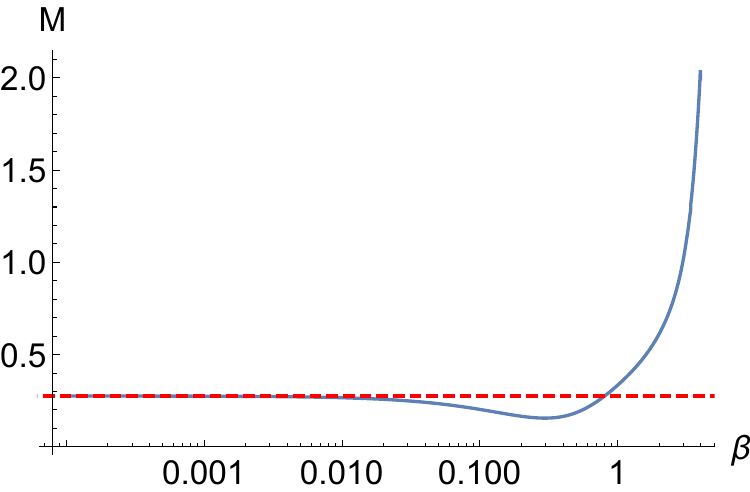}
	\caption{The scalar charge $D$, $\varphi_{\infty}$ and the BH mass $M$ as a function of $\beta$. The red dashed curves  denote the case with trivial scalar hair.}
	\label{case2-2}
\end{figure}

We also show the scalar charge $D$, $\varphi_{\infty}$ and the BH mass $M$ as a function of $\beta$ in Fig.\ref{case2-2}. It is evident that as $\beta$ increases, the scalar charge $D$ increases and eventually approaches infinity, while $\varphi_{\infty}$ monotonically decreases and tends towards negative infinity. Correspondingly, after a slightly decrease, the BH mass $M$ undergoes a significant rise with $\beta$ increasing, ultimately approaches infinity as well. This behavior arises because, in this limit, the coupling function $F(\varphi)$ approaches infinity, which is distinct from the case studied in the previous subsection. In addition, it is no doubt that the hair in this scenario is also secondary.

\subsection{The black hole thermodynamics} \label{case3}

The thermodynamic stability of these solutions is crucial for their physical plausibility. Therefore, it is necessary to investigate the thermodynamic characteristics of the hairy black holes that extends from the Schwarzschild solution and beyond.
Because we are working with the gravity theory involving higher derivative term, we adopt the entropy formula developed by Wald \cite{Wald:1993nt,Iyer:1994ys}
\begin{equation}
	S_h=2\pi \int_{\Sigma} d^2 x\sqrt{-h}\frac{\partial \mathcal{L}}{\partial R_{\mu\nu\rho\sigma}}\epsilon_{\mu\nu}\epsilon_{\rho\sigma}\,, \label{wald}
\end{equation}
where $\mathcal{L}$ represents the Lagrangian density, $\epsilon_{\mu\nu}$ is the binormal on the horizon $\Sigma$ and $h$ the induced metric. Specifically for our model, the BH entropy can be explicitly worked out as follows:
\begin{align}
	S_h=\pi-\frac{\pi \alpha F(\varphi (1))}{3 U_1(1)}\bigg[ & 3 U_2(1) U_1'(1)+U_1(1) (U_2'(1)\nonumber \\
	                                                         & +8 U_2(1)+4)\bigg]\,.
\end{align}
Once the entropy is at hand, the free energy $F_h$ can be determined by $F_h = M - T S_h$, where the temperature for our ansatz (\ref{ansatz}) is given by
\begin{equation}
	T=\frac{\sqrt{U_1(1) U_2(1)}}{4 \pi }\,.
\end{equation}

\begin{figure}[htbp]
	\centering
	\includegraphics[width=0.23\textwidth]{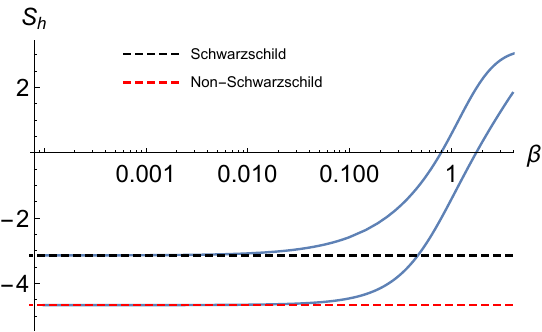}\hspace{2mm}
	\includegraphics[width=0.23\textwidth]{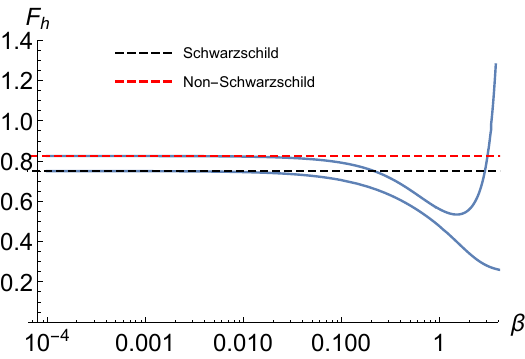}
	\caption{The entropy $S_h$ (left panel) and the free energy $F_h$ (right panel) of hairy BHs extending from the Schwarzschild and non-Schwarzschild solutions as a function of $\beta$. The black dashed line represents the Schwarzschild case, while the red dashed line represents the non-Schwarzschild case.} \label{SSFF}
\end{figure}

Fig.\ref{SSFF} shows the entropy $S_h$ (left panel) and the free energy $F_h$ (right panel) of hairy BH extending from the Schwarzschild and  the non-Schwarzschild solutions as functions of $\beta$. Notably, the entropy of the hairy BH extending from the non-Schwarzschild BH is consistently lower than that of the one extending from the Schwarzschild BH. Conversely, the free energy is always larger for the hairy BH extending from non-Schwarzschild BH. This feature indicates that the hairy BH extending from the Schwarzschild solution is thermodynamically favorable than the one extending from the non-Schwarzschild BH.

\section{Discussion}\label{conclusion}

In this paper, we present novel scalarized BH solutions beyond the GR framework. These scalarized BH solutions are extended from the Schwarzschild one and non-Schwarzschild one found in the pure Einstein-Weyl gravity \cite{Lu:2015cqa}.

Importantly, our direct solution of the full fourth-order EOMs was crucial for obtaining these novel solutions and uncovering new physics. By retaining the complete structure of the higher-derivative theory, we found solutions that differ significantly from those in the literature obtained from reduced second-order equations. In particular, our solutions go beyond the scalarized black holes found in 4-dimensional Einstein-scalar-Gauss-Bonnet theory \cite{Doneva:2017bvd,Silva:2017uqg,Antoniou:2017acq}, which are merely extensions of the Schwarzschild solutions in GR. We further studied the entropy and free energy of these new scalarized black holes. We find the solution extending from the Schwarzschild case exhibits higher entropy and lower free energy compared to the one extending from the non-Schwarzschild case. This indicates the Schwarzschild-extended solution is thermodynamically more stable.

In conclusion, by directly solving the full fourth-order equations of motion arising from the higher-derivative theory, our findings unveil the evasion of the no-hair theorem and discover new scalarized black hole solutions beyond those in GR. The significance of retaining the complete higher-derivative structure is also demonstrated.

In addition to the findings presented above, there are two important directions for further research on scalarization in higher derivatives that can enrich our understanding of this phenomenon. The first direction involves exploring more general higher derivatives and a richer style of coupling. This approach aims to expand and diversify the scenarios of scalarization. This direction offers the opportunity to explore a wider range of phenomena and mechanisms of scalarizations. The second direction is studying scalarization in higher derivatives using dynamical methods. This may involve solving the fourth-order equations dynamically. Although this poses significant challenges, tackling them will provide a solid foundation for the main findings and shed light on the potential imprints that these higher derivative theories may leave on GW.

\begin{acknowledgments}

We are very grateful to Yang Huang, Mujing Li, Ruihong Yue and Hongsheng Zhang for helpful discussions and suggestions. This work is supported by the National Key R\&D Program of China (Grant No. 2020YFC2201400), the Natural Science Foundation of China (Grant Nos. 11905083, 12375054 and 12375055), the Science and Technology Planning Project of Guangzhou (202201010655), the Postgraduate Research \& Practice Innovation Program of Jiangsu Province (Grant No. KYCX21\_3192), and the Natural Science Foundation of Jiangsu Province (Grant No. BK20211601).
	
\end{acknowledgments}

\bibliographystyle{apsrev4-2f}
\bibliography{Ref}

\newpage
\appendix
\section{Numerical methods for solving the fourth-order equations of motion}\label{appendix:methods}

Numerical methods play a crucial role in solving differential equations. In this section, we will display the application of numerical methods to solve second and fourth-order equations.

To begin, let us consider the equations for functions $f$, $h$, and $\varphi$, which are functions of the variable $z$. We discretize the axis $z$ using the Gauss-Lobatto quadrature. This discretization allows us to represent $U_1$, $U_2$, and $\varphi$ as vectors $\mathbf{W}$. Therefore, the problem can be reformulated as finding the appropriate values for the elements of $\mathbf{W}$ that satisfy the equations. To recover the function representation of the discretized values, a convenient approach is to employ the Fast Fourier Transform (FFT) method. By applying the FFT, we can obtain the expansion coefficients, denoted as $a_i$. These coefficients can subsequently be utilized to reconstruct the function using a combination of Chebyshev polynomials, denoted as $a_i T^i(z)$.

Derivatives of these functions up to fourth order can be obtained using standard methods, such as the `NDSolve' function in Mathematica. This capability allows us to accurately compute the derivatives required for our numerical methods.

Due to the highly nonlinearity of the equations, we employ the Newton-Raphson iteration method. To apply this method, we first linearize the equations and then iteratively update the solution until convergence is achieved. It is important to note that the success of the iteration process heavily depends on the initial solution guess. Therefore, providing a good initial guess for the solution is crucial to obtaining accurate results.

We have observed that for the second-order case, varying the initial seeds leads to multiple branches of solutions. This suggests that the system possesses unfixed degrees of freedom, likely resulting from an inappropriate second-order reduction. However, for the fourth-order equations, we have only obtained four branches of solutions: Schwarzschild, non-Schwarzschild, and their scalarized versions. Remarkably, after testing thousands of random initial seeds, all successful cases converged to these four branches.

Furthermore, we explored the relationship between the solutions of the second-order and fourth-order cases. We found that when introducing the second-order solutions into the fourth-order equations, they often fail to satisfy the latter. However, when we introduce the solutions obtained from the fourth-order equations into the second-order case, they satisfy the system. This provides strong evidence that the fourth-order equations single out the correct solution space from the spaces obtained by the second-order reduction.

In the next section, we will present more explicit examples and further support our observations and conclusions.

\section{The validation of the solutions obtained from the fourth-order equations}\label{appendix:validation}

In this section, we provide concrete numerical examples to demonstrate that the solutions derived from the fourth-order equations consistently satisfy the reduced second-order equation systems. Conversely, the solutions obtained from the second-order equations often fail to satisfy the fourth-order equations. Our results of this supplementary have been made available in an open-source repository (\href{https://github.com/physicsuniverse/Scalarized_Weyl}{Repo link}).

To effectively evaluate the accuracy of the solutions, we introduce the residual $\bm{\mathcal{R}}^I(z)$ of the I-th EOMs as a measure, which is obtained by discretizing the equations of motion,
\begin{equation}
	\bm{\mathcal{R}}^I(z) = \text{{\it discretized eoms}} \, \bigg|_{\mathbf{W}(z)=\mathbf{W}_N(z)} \,. \label{residualEq}
\end{equation}
Ideally, the residual function tends to zero for exact solutions, indicating that the numerical results precisely satisfy the EOMs. However, in the case of numerical solutions, we anticipate an approximate zero value, denoted as $\bm{\mathcal{R}}^I(z) \approx 0$. Since the complete second-order and fourth-order EOMs are too lengthy, we have included them in the provided repository for convenience, rather than presenting them directly in this document.

\begin{figure}[htbp]
	\centering
	\includegraphics[width=0.23\textwidth]{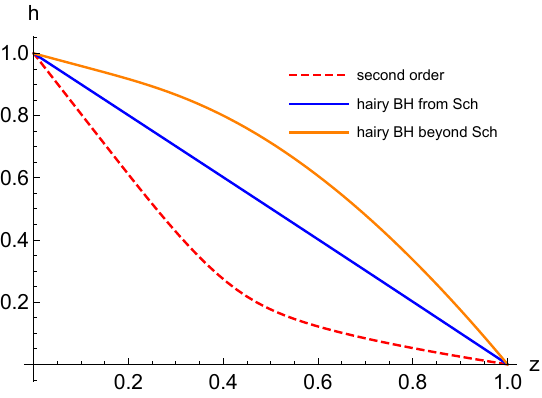}\hspace{2mm}
	\includegraphics[width=0.23\textwidth]{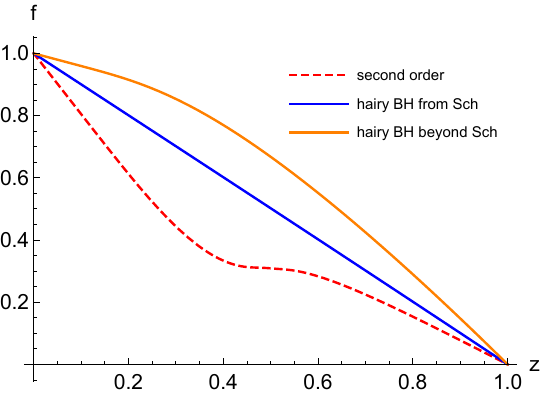}\hspace{2mm}
	\includegraphics[width=0.23\textwidth]{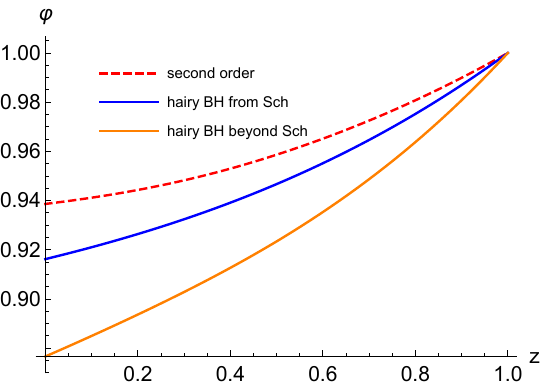}
	\caption{The metric fields $h$ and $f$, along with the scalar field $\varphi$ as functions of the radial coordinate $z$ for $\alpha=0.5$ and $\beta=0.1$. The red dashed lines are the branch of solutions solved by the reduced second-order equations. The blue and orange solid lines depict the solutions obtained from the original fourth-order equations, which have been discussed in the main text.}
	\label{Order2sol}
\end{figure}

\begin{figure}[htbp]
	\centering
	\includegraphics[width=0.23\textwidth]{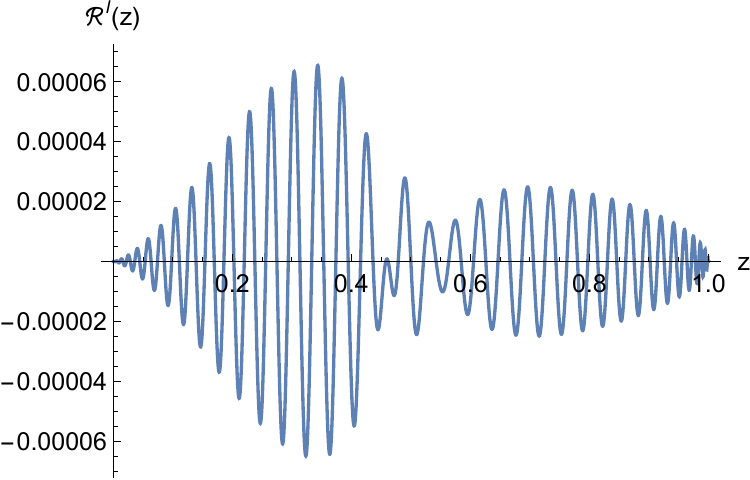}\hspace{2mm}
	\includegraphics[width=0.23\textwidth]{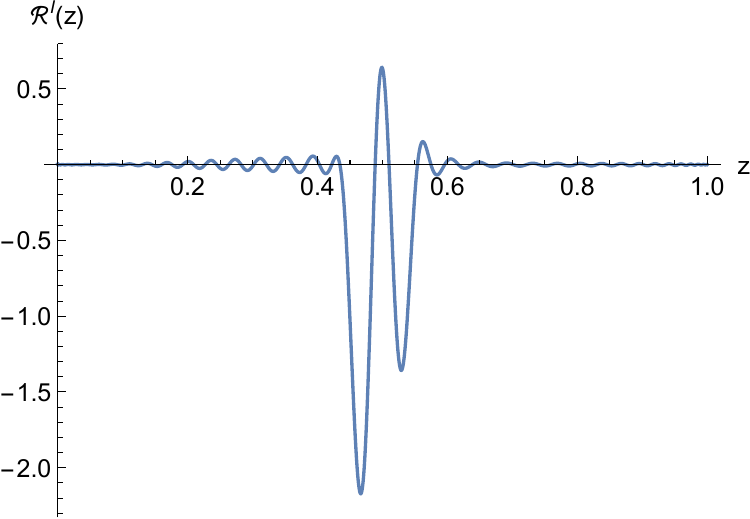}
	\caption{The residual function $\mathcal{R}^I(z)$ for the solutions obtained from the reduced second-order equations.  The left panel illustrates the scenario where the solution is brought back to the second-order equations, while the right panel demonstrates the case where the solution is brought back to the fourth-order equations.}
	\label{sol2ToEQ}
\end{figure}

As an example, let us set $\alpha=0.5$ and $\beta=0.1$. Numerical simulations reveal that the reduced second-order equations yield multiple solutions instead of just fixed ones. Fig.\ref{Order2sol} depicts one branch of these solutions with red dashed lines. Then, we evaluate the residual function for this solution. The left panel in Fig.\ref{sol2ToEQ} illustrates the one of the residual $\bm{\mathcal{R}}^I(z)$ where the solution is brought back to the second-order equations, while the right panel demonstrates the case where it is brought back to the fourth-order equations. It is evident that the residual function shown in the left panel of Fig.\ref{sol2ToEQ} is approximately zero. However, we observe that the residual function exhibited in the right panel significantly exceeds $\mathcal{O}(1)$. This indicates that the solution obtained from the reduced second-order equations fails to satisfy the original fourth-order equations.

\begin{figure}[htbp]
	\centering
	\includegraphics[width=0.23\textwidth]{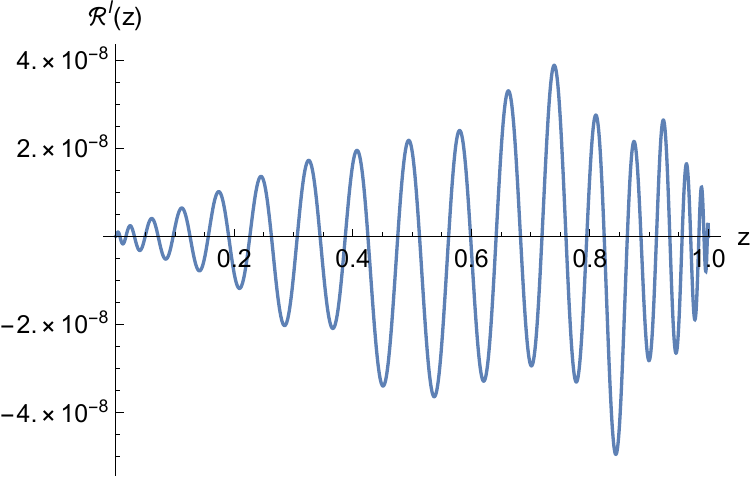}\hspace{2mm}
	\includegraphics[width=0.23\textwidth]{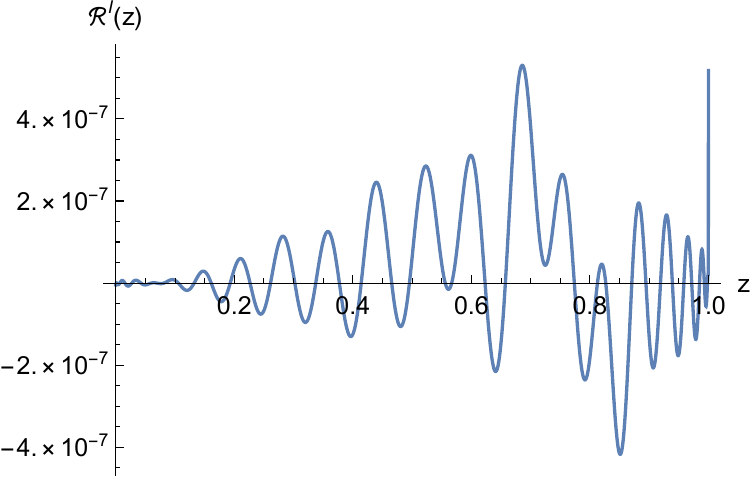}
	\caption{The residual function $\mathcal{R}^I(z)$ for the solutions obtained from the original fourth-order equations.  The left panel illustrates the scenario where the solution is brought back to the second-order equations, while the right panel demonstrates the case where the solution is brought back to the fourth-order equations.}
	\label{sol4ToEQ}
\end{figure}

However, regardless of whether we bring the solution obtained from the fourth-order equations back to the reduced second-order equations or the original fourth-order equations, we observe that the residual function is approximately zero, as depicted in Fig.\ref{sol4ToEQ}. This suggests that the solution satisfies both the second-order equations and the fourth-order equations.

\end{document}